# Fundamental defect of the macroeconomic thinking as one of the main causes of the crisis endured

Eugen Perchik

*Northeastern scientific center of NAS and MES of Ukraine (Kharkov, Ukraine)*
*Corporation "Scientific and technological institute of transcription, translation and replication" (Kharkov, Ukraine)*
E-mail address: eperchik@bk.ru

***Abstract.*** The main points of the first section of the article written by S.I. Chernyshov, A.V. Voronin and S.A. Razumovsky arXiv:1003.4382), which deals with the fundamental bases of the macroeconomic theory, have been analyzed. An incorrectness of the Harrod's model of the economical growth in its generally accepted interpretation was specifically considered. The inevitability of the economic crisis has been shown to follow directly from the premises of this model. At the same time there is an opportunity to realize the damping strategies.

In 2009 the Section the Social Science and Humanities of the NAS of Ukraine prepared a national report "Socio-economic condition of Ukraine: consequences for the people and state", 687 p., edited by V.M. Geits, A.I. Danilenko, M.G. Julinsky, E.M. Libanova, O.S. Onischenko. The range of macroeconomic problems is represented in chapter 3 "Influence of the world crisis on Ukraine and anti-crisis strategy of its socio-economic condition".

Referring to the international experts, the one of the causes of the world crisis was qualified: "an inconsistent macroeconomic policy which was mainly limited to the non-effective regulation of the financial markets, but wasn't aimed at the structural reforms". "An inclination of the economic agents to the excessive risk, which was provoked by the long period of the economic growth and macroeconomic stability" was also noticed.

It is worth emphasizing that the analysis of economical evolution, as well as the elaboration of regulatory decisions, is fundamentally based on the use of mathematical models. They are widely presented in the editions of various formats and to a considerable extent have shaped the current macroeconomic thinking. The conclusions and recommendations which follow from the mentioned models penetrated into the social mind and, reaching the level of mentality, became a driving force of its activity. First of all, this concerns the well-known Harrod's model which is illusively very simple.

Really, the yearly income $Y(\tau)$, $\tau = 1, 2, ...$ is divided into the volumes of consumption $C(\tau)$ and investment $I(\tau)$:

$$Y(\tau) = C(\tau) + I(\tau); \qquad (1)$$

with

$$I(\tau) = \mu Y(\tau), \; 0 < \mu < 1. \qquad (2)$$

The following relationships are also used:

$$d_\tau K(\tau) = I(\tau), \qquad (3)$$

where $K(\tau)$ – is the capital; $d_\tau = d/d\tau$, $\tau$ – the dimensionless time, and

$$K(\tau) = \nu Y(\tau), \qquad (4)$$

where $\nu$ is the number of years during which the income, as it is used to say, "counterpoise" the capital.

Relationships (2) – (4) lead to a differential equation

$$d_\tau K(\tau) - \sigma K(\tau) = 0, \; \sigma = \mu/\nu, \qquad (5)$$

which solution is

$$K(\tau) = K_0 e^{\sigma \tau}, \; K_0 = K(0) \qquad (6)$$

and accordingly

$$Y(\tau) = Y_0 e^{\sigma \tau}; \; I(\tau) = I_0 e^{\sigma \tau}, \; Y_0 = Y(0); \; I_0 = I(0), \qquad (7)$$

where $Y_0 = K_0/\nu$; $I_0 = \sigma K_0$.

Expressions (6), (7) argues that the capital, income and investment interactively (in other words, their diagrams are not crossed) growth exponentially during the unlimited period of time. This kind of economics has no serious problems because:

- can be forecasted and is stable with respect to perturbations of its growth which are synchronically compensated balancing the system, in fact due to peculiarities of the solution of the differential equation with the constant coefficient (5);

- acceleration or, on the contrary, deceleration of this growth is determined by the parameters $\mu$ and $\nu$, which possess definite meaning and in these terms the economics is manageable;

- realization of the relationship (4) is a guaranty of so favorable dynamics, in other words, in the year $j$ the received income must be $K(\tau_j)/\nu$, $\tau_j = j$; $j = 0, 1, 2, \ldots$ .

The deep analysis of the Harrod's model (1) – (4) was presented in the first section of the article written by S.I. Chernyshov, A.V. Voronin and S.A. Rasumovkky "The problem of modelling of economic dynamics" (arXiv:1003.4382; www.ttr.com.ua; http://chvr-article.narod.ru, further – chvr).

Proceeding from the difficult-to-forecast consequences of the differentiation in the relationship (3), connected with the appearance of $\delta$-functions:

$$d_\tau K(\tau) = \sum_{j=0}^{n} I_j \delta(\tau - \tau_j), \; \delta(\tau) = \begin{cases} 0, \tau \neq 0; \\ \infty, \tau = 0, \end{cases} \int_{-\infty}^{\infty} \delta(\eta) d\eta = 1, \qquad (8)$$

the authors used the law of the capital forming [chvr]

$$K_n = K_0 + \sum_{j=1}^{n} I_j, \qquad (9)$$

where $K_n = K(n)$; $I_j = I(j)$. This sum, strictly speaking, is a base for getting (3).

Since relationships (1), (2), and (4) are discrete they can be presented in the form similar to the (9):

$$Y_n = C_n + I_n; \; I_n = \mu Y_n; \qquad (10)$$

$$K_n = \nu Y_n, \tag{11}$$

where $Y_n = Y(n); C_n = C(n)$.

The relationships (9) – (11) embodies the Harrod's model in the difference (or discrete) interpretation. It flows that

$$K_n = K_0 e^{n\ln(1+\sigma)}; Y_n = Y_0 e^{n\ln(1+\sigma)}; I_n = I_0 e^{n\ln(1+\sigma)}, \tag{12}$$

and since the quantity $\sigma$ is actually rather small (we can admit $\mu \sim 0,5$; $\nu \sim 10$) and accordingly

$$\ln(1+\sigma) \approx \sigma, \tag{13}$$

then using this replacement we get solution

$$K_n = K_0 e^{\sigma n}; Y_n = Y_0 e^{\sigma n}; I_n = I_0 e^{\sigma n}, \tag{14}$$

which is identical to the expressions (6), (7) when $\tau = n = 0, 1, 2, \dots$.

So, it seems possible to conclude that the paradigm of the stable growth of the economy for the unlimited perspective according to (6), (7) is undoubted. This paradigm was deeply rooted during many decades and is supported by the authority of classic. In practical respect it means that economic crises are casual; they can be prevented by strictly following exponential dependencies (6), (7), and accordingly, relationship (4). In any case, "an ideal" is exist and is reachable in essence.

Within this logic the situation when an income $Y_n$, composes a part $\nu$ of the capital $K_n$, was received less then due in the year $n$ (in other words, the relationship (11) was violated), is just the cause of the poor economic situation. Accordingly, to eliminate the lagging, an income $Y_{n+1}$ in the year $n+1$ must be greater then $\nu$ part of the capital $K_{n+1}$. It is possible to simultaneously manipulate the level of consumption $\tilde{N}_n(\tau)$.

At the same time, a peculiar point is emphasised in [chvr]. For the exact solution (12) when the value of $n$ in (9) is close to $\sigma^{-1}$, the capital which was formed during the period starting from the first year, namely

$$\sum_{j=1}^{n} I_j = \sigma K_0 e^{\ln(1+\sigma)} \frac{e^{n\ln(1+\sigma)} - 1}{e^{\ln(1+\sigma)} - 1}$$

(formula for the sum of geometric progression was used), appears to be equal to the whole capital $K_n$. This only possible when $K_0 = 0$, and so we face a contradiction.

If "an approximate" (because of using (13)) solution (14) is substituted into expression for $K_n$ (9) instead of (12) the, the contradiction is eliminated immediately. Note that the peculiarity of the solution when $n$ is close to $\sigma^{-1}$ doesn't follow directly from expression (12).

It follows that the premise (13) is far from being inoffensive. Analogously the correctness of using the derivative of the capital $K(\tau)$, understood in its usual meaning (3) instead of generalized function (8), becomes questionable. The essence is that the relationship (3) is inherent in deriving the differential equation (5), which has a solution (6), (7), coinciding with (14). In other words, in deducing this equation the discrete function is treated as the continuous one.

The above considerations brought [chvr] to the conclusion that categories of the continuous analysis need to be used consistently (in contrast to the symbiosis (1) – (4)). Dimensionless time $\tau$ is linked to the same unit – 1 year, but unlike the discrete case it can be subdivided into arbitrary small intervals. In this way the question about the correctness of the relationship (3) is removed. Relationships (1) and (2) are also directly extended to the case of continuous argument $\tau$.

From (3) it follows that

$$K(\tau) = \int_{-T}^{\tau} I(\eta) d\eta = K_0 + K_R, \tag{15}$$

where $T$ is the period of the accumulation of the initial capital $K_0$;

$$K_R(\tau) = \int_{0}^{\tau} I(\eta) d\eta \tag{16}$$

is the capital realized over the period $\tau > 0$.

Generally speaking, a continuous analog of the sum (9) have been obtained, but now $I(\tau)$ is the intensity of flow of investments measured in the money equivalent normalized to the dimensionless time. In virtue of (1) and (2), $Y(\tau)$ and $C(\tau)$ are also intensities of flow of income and consumption accordingly.

Along with this, the remaining relationship (4), or (11), is essentially discrete because for the continuous argument $\tau$:

– the discrepancy of the dimensionalities takes place, namely, the capital in the money equivalent is compared through the dimensionless coefficient $\nu$ with the income measured in the money equivalent normalised to the dimensionless time;

– when $\tau$ is close to 0, the value of $\nu$ tends to ∞ and accordingly an uncertainty appears (really, to counterbalance the capital with an income received, for example, in 1 minute we need a crowd of such minutes).

The authors of [chvr] cogently demonstrated that the analog of (4) in the case of dimensionless time $\tau$ is the relationship

$$K(\tau) = \frac{\nu}{\tau}\int_0^\tau Y(\eta)d\eta; \qquad (17)$$

its peculiarity at $\tau = 0$ is eliminated according to the L'Hospital rule reshaping into initial identity $K_0 = \nu Y_0$. This relationship is essentially based on the premise that the capital must be compared with the income for some period of time (accordingly an integral of its intensity emerges, which is measured in the money equivalent).

So the continuous Harrod's model, which is the subject of the further part of the investigation, is defined by relationships (17) and (1) – (3), namely

$$Y(\tau) = C(\tau) + I(\tau); \quad I(\tau) = \mu Y(\tau); \quad d_\tau K(\tau) = I(\tau). \qquad (18)$$

where $Y(\tau)$; $C(\tau)$ and $I(\tau)$ are the intensities of the financial flows with the dimension money equivalent/unit of the dimensionless time.

This model generates solution

$$K(\tau) = \frac{K_0}{1-\sigma\tau}; \quad I(\tau) = \frac{I_0}{(1-\sigma\tau)^2}; \quad Y(\tau) = \frac{1}{\mu}I(\tau); \quad C(\tau) = (1-\mu)Y(\tau), \quad (19)$$

which corresponds with a Cauchy problem:

$$d_\tau K(\tau) - \frac{\sigma}{1-\sigma\tau}K(\tau) = 0, \quad K(0) = K_0;$$

$$d_\tau I(\tau) - \frac{2\sigma}{1-\sigma\tau}I(\tau) = 0, \quad I(0) = \sigma K_0 \qquad (20)$$

and, obviously, at $\tau = \sigma^{-1}$ there is a peculiarity, which is in agreement with the result obtained for the discrete time $\tau$ in the statement (9) – (11).

Note the following aspects:

- unlike the discrete case (12), the solution (19) represents the mentioned peculiarity explicitly;

- this reflects essential distinctions between differential and difference models (which are not properly investigated yet);

- equations (20) possess variable coefficients and thus fundamentally differ from (5).

It is noticed [chvr] that at the very beginning when $\tau = v$ an income realized over the period of time $0$ to $\tau$

$$Y_R(\tau) = \int_0^\tau Y(\eta)d\eta,$$

according to (17) becomes equal to the capital $K(v)$. Later, when time $\tau = \sigma^{-1}$ is reached relationship (17), with the account of $Y(\tau) = I(\tau)/\mu$, gets the form

$$K(\sigma^{-1}) = \int_0^{\sigma^{-1}} I(\eta)d\eta$$

and from the comparison with (15), (16) it follows that

$$K(\sigma^{-1}) = K_R(\sigma^{-1}); \quad K_0 = 0.$$

Meanwhile an initial capital $K_0$ instantly overflows into an income $Y_R(\sigma^{-1})$ and expressions (19) transform into uncertainty. Accordingly, Cauchy problem (20) loses its sense when $K_0 = 0$. This situation can be interpreted from the standpoint of the economic chaos and collapse origin and also from the standpoint of the limitedness of the look-ahead period (mentioned phenomena could become a subject of a separate investigation).

Note that to define functions which represent the continuous Harrod's model (17), (18), we need only three equations which interconnect the capital $K(\tau)$, income $Y(\tau)$ and investments $I(\tau)$. There are equations (17) and (2), (3), included in (18). From the remaining equation (1) the intensity of the consumption flow $C(\tau)$ can be found according to the calculation of an income $Y(\tau)$ and investments $I(\tau)$. Therefore, from the mathematical point of view an increment of the function $C(\tau)$ has no direct influence on the economics.

It follows that starting from some time $\tau < \sigma^{-1}$, it is reasonable to transfer (passing an intermediate chain $Y_R(\tau)$) the initial capital $K_0$ into consumption

$$C_R(\tau) = \int_0^\tau C(\eta) d\eta$$

and this seems to be economically logical. Really, the economic development must be mainly innovative and outdated capital must be taken out of the industrial and technological sphere in proper time.

In practice the beginning of this process can be presented at the level $\tau = \nu$, when in virtue of (15) – (17)

$$K(\nu) = Y_R(\nu) = K_0 + K_R(\nu)$$

and since $I_R(\tau) = K_R(\tau)$, where

$$I_R(\tau) = \int_0^\tau I(\eta) d\eta$$

is the volume of the investments realized over a period of time $0$ to $\tau$, with the account of (1), we get $C_R(\nu) = K_0$.

This seems to have a conceptual interpretation. Really, when the consumption reaches the value of the initial capital it is quite a reasonable criterion for the reformatting the economics, including the structural innovations, with the transition to the next level of development. In other words, the initial capital has been "wasted" and it is time to do something.

Further it is noticed in [chvr] that parameter $v$ is less defined in prognostic and managerial sense then $\mu$ because it depends on industrial factors, subject to the influence of conjuncture within particular time interval and so on. Therefore it is interesting to define it more exactly on the sign boundary of the instant of time $v$ using conditions

$$K(v) = K_0/(1-\mu); \quad I(v) = I_0/(1-\mu)^2; \quad Y(v) = Y_0/(1-\mu)^2,$$

which are obtained if we substitute $\tau = v$ in (19).

In other words, if, for example, $Y(\tau)$ coincides with the value of $Y_0/(1-\mu^2)$, we can judge about the value of $v$. Thus there is the possibility to represent realities of the situation under analysis and to correct economic strategies.

The generalization of the relationship (17) is presented for the same purpose:

$$K(\tau) = \frac{v}{f(\tau)} \int_0^\tau Y(\eta) d\eta, \qquad (21)$$

where the function $f(\tau)$ satisfies the condition of elimination of the peculiarity in zero $d_\tau f(0) = 1$. In particular, this may be the polynomial of the form

$$f(\tau) = \tau + \sum_{n=2}^N a_n \tau^n.$$

Using (18), (21), we get

$$K(\tau) = \frac{K_0}{1 - \sigma f(\tau)}$$

and it is obvious that the crisis corresponding to the solution of the equation $f(\tau) = \sigma^{-1}$, can only be postponed by decelerating the capital accumulation. The function $f(\tau)$ determined proceeding from the growth of $K(\tau)$ within representative enough time interval, say $[0, \nu]$, can be extrapolated to the critical value of $\tau$.

Thus "the crisis" is inherent not only in the Harrod'd model in the statement (17), (18), and this is the very important conclusion. It emerges as a consequence of practically any dependence between the capital $K(\tau)$ and income over the period of time $Y_R(\tau)$.

It is worth emphasizing that conceptual premises of the Harrod's model, correctly interpreted, bring us to the solution, according to which economic crises are inevitable. The illusion of the exponential growth (6), (7) and (14) is a consequence of the inadequacy of the mathematic model (1) – (4). It is likely because discoverers made a fetish from the final result, prompted by their investigations in the categories of economics (ignoring the fact of inaccurate treatment of infinitesimals). But this mistake results in the enormous strata of biased economic regulations, conclusions and recommendation which still be using in managerial decisions.

To develop the Harrod's model, the Cauchy problem has been formulated in [chvr] for the cases of time dependent parameter $\mu$ in (18):

$$d_\tau K(\tau) - \frac{\mu(\tau)}{\nu - \tau\mu(\tau)} K(\tau) = 0, \; K(0) = K_0$$

and for the accounting of the amortization by means of substitution of the expression $(1-\alpha\tau)K(\tau)$ instead of $K(\tau)$ in (17), (18), where $\alpha > 0$ is a constant:

$$d_\tau K(\tau) - \frac{\alpha + \sigma - 2\alpha\sigma\tau}{1 - (\alpha+\sigma)\tau + \alpha\sigma\tau^2} K(\tau) = 0, \; K(0) = K_0.$$

The final step of the first section of the considered work by S.Y. Chernyshov and so-authors is the development of the Harrod's model for the case of the cumulative effect when investments flowing into the capital, which implies the scientific and technical progress, market price and other factors. It is represented by using the presentation

$$K(\tau) = \int_{-T}^{\tau} (1+\rho\eta) I(\eta) d\eta,$$

instead of (15), where $\rho > 0$ is a constant. In other words, with time the value of the investment is transforming into the capital growth faster then the simple addition.

The problem is reduced to the differential equation

$$d_\tau K(\tau) - \frac{\sigma(1+\rho\tau)}{1-\sigma\tau-\sigma\rho\tau^2} K(\tau) = 0,$$

whose solution

$$K(\tau) = K_0 \left\{ \frac{1}{\sqrt{-\sigma\rho\tau^2 - \sigma\tau + 1}} + \left[ \frac{-2\sigma\tau - \sigma - \sqrt{4\sigma\rho + \sigma^2}}{-2\sigma\tau - \sigma + \sqrt{4\sigma\rho + \sigma^2}} \right]^{\frac{\sigma}{2\sqrt{4\sigma\rho+\sigma^2}}} \right\}$$

becomes $\infty$ (to be more exact, there is an uncertainty similar to (19)) when

$$\tau = -\frac{1}{2\rho} + \sqrt{\frac{1}{4\rho^2} + \frac{1}{\sigma\rho}}$$

and, obviously, the increase of $\rho$ hasten the crisis.

From our point of view, results of the work by S.Y. Chernyshov and so-authors may be qualified from the standpoint of the newly formed paradigm of the world macroeconomic thought. Really, it was pronouncedly demonstrated for the first time that apparently spontaneous crisises in fact are not anomalies. In opposite, they are organically inherent in the development of the economic system.

The long-term stable growth of the economic does not confirmed by the practice, but nevertheless the belief that this "ideal" is attainable have firmly seized the social mind. Most likely this is the result of natural hopes of any human being for the better life and prosperity. These hopes were scientifically supported by the theory of exponential growth, which require just a reasonable balance between the capital and income.

The external convincingness of the arguments of this theory was reinforced by using the differential equation (framed with the macroeconomic rhetoric) which was demonstrated to be incorrect by S.I. Chernyshov, A.V. Voronin and S.A. Razumovsky.

At the same time it must be stated that the above proportion is the only instrument suggested to the society for regulating the economics.

From this point of view the methodology of S.I. Chernyshov and co-authors seems to be rather optimistic. Really, crisis is inevitable and is hastened by the activization of the economics that includes the scientific and technical progress, but there are effective means to prevent the extreme situation. The following are among them:

- the possibility a priori estimate pre-crisis processes for identification of the instant of its origin and undertaking preventive measures;

- the use of the initial capital as the instrument for the smooth ending of the phase of economic development by means of its adaptive amortization and transfer into the sphere of consumption.

Note in conclusion that in second section of the work by S.I. Chernyshov and so-authors "The problem of modeling of economic dynamics" a more general Phillips model is examined. The defect of the symbiosis of the continuous and discrete analysis, which is analogous to the Harrod's model, has been also eliminated. The problem is reduced to the solution of the ordinary differential equation of the second order with variable coefficients. As a result of the inadequate use of the notion of infinitesimal quantity, the other models of macroeconomic dynamics in the differential form (Harrod-Domar's, Gudvin's, Kaletsky's) appeared to be mistaken.